\documentstyle[aps,preprint]{revtex}

\begin{document}
\draft
\title{On the compressibility equation of state\\
for multicomponent adhesive hard sphere fluids.}
\author{Domenico Gazzillo and Achille Giacometti}
\address{Istituto Nazionale di Fisica della Materia and \\
Dipartimento di Chimica Fisica, 
Facolt\`{a} di Scienze, Universit\`{a} di Venezia, \\
S. Marta DD 2137, I-30123 Venezia, Italy}
\date{\today }
\maketitle

\begin{abstract}
The compressibility equation of state for a multicomponent fluid
of particles interacting via an infinitely narrow and deep potential,
is considered within the mean spherical approximation (MSA).
It is shown that for a class of models leading to a particular form
of the Baxter functions $q_{ij}(r)$ containing density-independent stickiness
coefficient, the compressibility EOS does not
exist, unlike the one-component case. The reason for this is that
a direct integration of the compressibility at fixed composition, cannot
be carried out due to the lack of a reciprocity relation on the second
order partial derivatives of the pressure with respect to two different
densities. This is, in turn, related to the inadequacy of the MSA.
A way out to this drawback is presented in a particular
example, leading to a consistent compressibility pressure, and a possible
generalization of this result is discussed.
\end{abstract}

\newpage

\section{INTRODUCTION}

Baxter's `sticky hard sphere' model \cite{Baxter68,Perram75} (hereafter
referred to as {\it SHS1 model}) has often been employed in studies on
colloidal suspensions of adhesive particles. Its potential adds to a hard
sphere (HS) repulsion an infinitely strong surface adhesion, defined by
taking an attractive square-well tail with vanishing width and infinitely
increasing depth, giving a finite non-zero contribution to the second virial
coefficient (`sticky limit') \cite{Baxter68}. The SHS1 model admits
analytical solution if the Ornstein-Zernike (OZ) integral equations of the
liquid state theory are coupled with the Percus-Yevick (PY) approximation %
\cite{Baxter68,Perram75,Barboy79}. The resulting expression for a fluid with 
$p$ components requires the knowledge of a set of density-dependent
parameters $\left\{ \lambda _{ij}\right\} $, whose values have to be
determined by solving numerically $p(p+1)/2$ coupled quadratic equations %
\cite{Perram75}. The applicability of the SHS1-PY solution is, therefore,
limited to systems with a small number of components \cite{Robertus89}.

On the other hand, colloidal suspensions are rather commonly polydisperse.
Polydispersity means that mesoscopic suspended particles of a same chemical
species are not necessarily identical, but some of their properties (size,
charge, etc.) may exhibit a discrete or continuous distribution of values.
Even when all macroparticles belong to a unique chemical species, a
polydisperse fluid must therefore be treated as a multicomponent mixture,
with very large $p$ values - of order $10^{1}\div 10^{3}$ or more (discrete
polydispersity) - or with $p\rightarrow \infty $ (continuous polydispersity).

The above-mentioned shortcomings of the SHS1-PY solution offer a strong
motivation for investigating an alternative sticky hard sphere model,
proposed by Brey {\it et al.} \cite{Brey87} and Mier-y-Teran {\it et al.} %
\cite{Mier89}, and hereafter referred to as {\it SHS2 model}. The adhesive
part of its potential is defined starting from an attractive Yukawa tail,
which, in the sticky limit, has both amplitude and inverse range tending to
infinity, with their ratio remaining constant. In this case, the OZ
equations are analytically solvable within the {\it mean spherical
approximation} (MSA) \cite{Mier89,Ginoza96}. Although the SHS2-MSA solution
is simpler than the SHS1-PY one, it has received much less attention,
especially in the multi-component case.

In two previous papers \cite{Gazzillo00,Gazzillo02} we investigated
structural properties of polydisperse fluids using a version of the SHS2
model in which the coupling (stickiness) parameters, which define the
strength of the Yukawa attraction, are {\it factorizable}. This choice is
the simplest one \cite{Yasutomi96,Herrera98}. In fact, a slight different
version of the SHS2 potential (with non-factorizable coefficients), proposed
by Tutschka and Kahl \cite{Tutschka98,Tutschka00,Tutschka01} leads to more
complicated analytical results, without any increase of the physical insight.

Once that the structural properties of a model are known, the next natural
step is to study the corresponding thermodynamics. Unfortunately, neither
for SHS1 \cite{Barboy79} nor for SHS2 (see below), this is a simple task.

In this short contribution we focus in particular on the compressibility
equation of state (EOS) of the SHS2 multi-component model, since the
compressibility route represents the simplest method for obtaining the
pressure, given an analytical expression for Baxter's factor correlation
function $q_{ij}(r)$. Under rather general conditions, we show that no
compressibility EOS can exist for the SHS2 multi-component model, within the
MSA. We argue how this inconsistency stems from the MSA closure, and show a
possible way out to overcome this difficulty, by using a simple illustrative
example.

After this work has been completed, we became aware of Ginoza's recent
analysis \cite{Ginoza01}, where the author discusses a factorizable model
essentially identical to the one considered in our previous work \cite%
{Gazzillo00}, but fails to recognize the MSA inconsistency, and thus reports
an incorrect expression for the compressibility EOS.

Our findings agree with those by Tutschka and Kahl \cite%
{Tutschka00,Tutschka01}, who observed the same inconsistency within their
particular version of the SHS2 model.

\section{THE SHS2 MODEL}

The starting point of the SHS2 model is a fluid with particles interacting
via a HS repulsion plus a particular Yukawa (HSY) attraction, i.e.%
\begin{equation}
u_{ij}^{{\rm HSY}}(r)=\left\{ 
\begin{array}{lll}
+\infty , &  & 0<r<\sigma _{ij}=(\sigma _{i}+\sigma _{j})/2 \\ 
-zA_{ij}e^{-z(r-\sigma _{ij})}/r, &  & r\geq \sigma _{ij}.%
\end{array}%
\right.  \label{s1}
\end{equation}

\noindent \noindent Here, $\sigma _{i}$ denotes the HS diameter of species $%
i $ (whose number density is $\rho _{i}$), $z$ is the inverse range of the
Yukawa tail, all $A_{ij}=A_{ji}$ stickiness parameters are $\geq 0$, and the
well depth, $\varepsilon _{ij}=zA_{ij}/\sigma _{ij}$, depends on $z$
linearly. The OZ equations for HSY mixtures have been solved analytically %
\cite{Blum78}, for any finite $z,$ within the MSA closure, i.e., $%
c_{ij}\left( r\right) =-\beta u_{ij}(r)$ for $r>\sigma _{ij}$ ($c_{ij}\left(
r\right) $ is the direct correlation function, and $\beta =(k_{B}T)^{-1}$).
The MSA solution for SHS2 can thus be obtained by taking the sticky limit, $%
z\rightarrow \infty $, of the solution $q_{ij}^{{\rm HSY-MSA}}(r)$ for the
HSY fluid. The result is%
\begin{equation}
q_{ij}(r)=\left\{ 
\begin{array}{l}
\frac{1}{2}a_{i}(r^{2}-\sigma _{ij}^{2})+b_{i}(r-\sigma _{ij})+K_{ij},\qquad
L_{ij}\leq r\leq \sigma _{ij} \\ 
0,\qquad \qquad \qquad \qquad \text{elsewhere}%
\end{array}%
\right.  \label{s2}
\end{equation}%
\begin{equation}
a_{i}=\frac{1}{\Delta }+\frac{3\xi _{2}\sigma _{i}}{\Delta ^{2}}-\frac{X_{i}%
}{\Delta },\qquad b_{i}=-\frac{3\xi _{2}\sigma _{i}^{2}}{2\Delta ^{2}}+\frac{%
X_{i}\sigma _{i}}{2\Delta },  \label{s3}
\end{equation}

\begin{equation}
X_{i}=\frac{\pi }{6}\sum_{m}\rho _{m}\sigma _{m}\ M_{im},\qquad
M_{im}=12K_{im},  \label{s4}
\end{equation}

\noindent where $L_{ij}=(\sigma _{i}-\sigma _{j})/2$, $\ $ $\xi _{n}=(\pi
/6)\sum_{l}\rho _{l}\sigma _{l}^{n},$ and $\Delta =1-\xi _{3}.$ The
coefficients $K_{ij}=q_{ij}(\sigma _{ij}^{-})$, given by

\begin{equation}
K_{ij}^{{\rm SHS2-MSA}}=\frac{A_{ij}}{k_{B}T}\ ,  \label{s5}
\end{equation}%
are {\it density-independent}, and have dimensions of [length]$^{2}.$

Tutschka and Kahl's version of the SHS2 model \cite%
{Tutschka98,Tutschka00,Tutschka01} hinges upon non-factorizable parameters, $%
K_{ij}^{{\rm SHS2-MSA}}=\gamma _{ij}\sigma _{ij}^{2}$, with $\gamma _{ij}$
obeying a Berthelot-type rule, i.e., $\gamma _{ij}=\left( \gamma _{ii}\gamma
_{jj}\right) ^{1/2}.$ Our version with {\it factorized} coefficients assumes
that $A_{ij}=\varepsilon _{0}G_{i}G_{j}$, where $\varepsilon _{0}$ is an
energy and $G_{m\text{ }}$has dimensions of length. Thus

\begin{equation}
K_{ij}=Y_{i}Y_{j}\ ,  \label{s6}
\end{equation}

\begin{equation}
\text{where}\qquad Y_{m}=\gamma _{0}G_{m},\qquad \ \gamma _{0}^{2}\ =\frac{%
\varepsilon _{0}}{k_{B}T}=\frac{1}{12T^{\ast }}\ ,  \label{s7}
\end{equation}%
with $T^{\ast }$ being a reduced temperature (as in Baxter's model, the
factor 1/12 is introduced for later convenience. In our previous papers \cite%
{Gazzillo00,Gazzillo02} it was absent, and the correspondence between the
two reduced temperatures is: $T^{\ast }=T_{{\rm old}}^{\ast }/12).$

Irrespectively of the choice for the coefficients $A_{ij}$, the SHS1 and
SHS2 potentials are {\it different} and should not be confused even in the
sticky limit. An intuitive way of understanding this point is to notice
that, as the well width goes to zero, the area of the square well in
Baxter's SHS1 model vanishes, whereas the area under the Yukawa tail in SHS2
remains finite \cite{Brey87}. This difference becomes important when
evaluating thermodynamics. In fact,\ it can be shown that the virial
pressure depends not only on the $q_{ij}(r)$ resulting after the sticky
limit, but also on the functional form of the tail as well as on the way the
sticky limit is taken \cite{Gazzillo02b}. Furthermore, the SHS1 model is
analytically solvable within the PY closure, but not within the MSA one,
whereas the opposite is true for SHS2.

At the level of approximate solution for $q_{ij}(r)$, the difference between
the SHS2-MSA and SHS1-PY expressions lies only in the coefficients $K_{ij},$
which in the latter case read

\begin{equation}
K_{ij}^{{\rm SHS1-PY}}=\frac{1}{12\tau _{ij}}y_{ij}^{{\rm PY}}(\sigma
_{ij})\sigma _{ij}^{2}\equiv \frac{1}{12}\lambda _{ij}\sigma _{ij}^{2},
\label{s9}
\end{equation}%
where the dimensionless positive parameter $\tau _{ij},$ that appears in the
SHS1 potential, is related to both the temperature and the stickiness
between particles of species $i$ and $j$, while $y_{ij}(\sigma _{ij})$ is
the contact value of the cavity function. Note that $K_{ij}^{{\rm SHS1-PY}}$
is non-factorizable and {\it density-dependent, }since $y_{ij}(\sigma _{ij})$
depends on the densities of all components in the mixture. This difference,
albeit seemingly harmless, has far reaching consequences, as it will be
shown in the following.

\section{COMPRESSIBILITY EQUATION OF STATE}

Once $q_{ij}(r)$ is known, one can calculate derivatives of the
compressibility ($c$) pressure by means of two general relations, obtained
from fluctuation theory in the grand-canonical ensemble and from Baxter's
factorization of the three-dimensional Fourier transform of $c_{ij}\left(
r\right) $, i.e.,

\begin{equation}
\left( \frac{\partial \beta P}{\partial \rho _{i}}\right) _{T,\rho
_{k}}=a_{i}-2\pi \sum_{m}\rho _{m}a_{m}\widehat{q}_{mi}(0),  \label{b1}
\end{equation}

\begin{equation}
\chi _{T}^{-1}=\left( \frac{\partial \beta P}{\partial \rho }\right) _{T,%
{\bf x}}=\sum_{i}x_{i}\left( \frac{\partial \beta P}{\partial \rho _{i}}%
\right) _{T,\rho _{k}}=\sum_{m}x_{i}a_{i}^{2}\ ,  \label{b2}
\end{equation}

\noindent \noindent where $\widehat{q}_{ij}\left( k\right) $ is the
one-dimensional Fourier transform of $q_{ij}\left( r\right) $, while $\chi
_{T}=\rho k_{B}TK_{T}$ denotes the isothermal susceptibility ($K_{T}$ being
the isothermal compressibility), and $a_{i}=1-2\pi \sum_{l}\rho _{l}\widehat{%
q}_{il}(0).$ Note that the pressure is a function of $(T,\rho _{1},\ldots
,\rho _{p})$ in Eq. ($\ref{b1}$), and of $(T,\rho ,{\bf x})$ in Eq. ($\ref%
{b2}$), where ${\bf x}=(x_{1},\ldots ,x_{p-1})$ represents the composition
in terms of molar fractions.

We can apply the previous relations to SHS models for colloidal fluids,
after observing that Eq. ($\ref{s2}$) yields $\widehat{q}%
_{mi}(0)=12^{-1}(a_{m}\sigma _{i}^{3}\ +3\Delta ^{-1}\sigma _{m}\sigma
_{i}^{2}-M_{mi}\sigma _{i})$. Inserting this term and the $a_{i}$ given by
Eq. ($\ref{s3}$) into Eqs. ($\ref{b1}$)-($\ref{b2}$), one gets an expression
for $\left( \partial \beta P/\partial \rho _{i}\right) _{T,\rho _{k}}$ and

\begin{equation}
\left( \frac{\partial \beta P}{\partial \rho }\right) _{T,{\bf x}}=\left( 
\frac{\partial \beta P}{\partial \rho }\right) _{T,{\bf x}}^{{\rm HS-PY}c}-%
\frac{2\left\langle X\right\rangle }{\Delta ^{2}}-\frac{6\xi
_{2}\left\langle \sigma X\right\rangle }{\Delta ^{3}}+\frac{\left\langle
X^{2}\right\rangle }{\Delta ^{2}},  \label{b3}
\end{equation}%
where we have introduced compositional averages, such as $\left\langle
f\right\rangle \equiv \sum_{m}x_{m}f_{m}$ and $\left\langle fg\right\rangle
\equiv \sum_{m}x_{m}f_{m}g_{m},$ and the derivative with superscript HS-PY$c$
refers to the compressibility pressure of the corresponding HS mixture,
evaluated within the PY approximation. Note that the two results hold for
both the SHS2-MSA and SHS1-PY solutions, if the corresponding $K_{ij}$'s are
used.

In order to obtain the compressibility pressure, it might now be spontaneous
to perform immediately the integration of Eq. ($\ref{b3}$) with respect to $%
\rho ,$ at fixed composition ${\bf x},$ i.e. $\beta P(T,\rho ,{\bf x}%
)=\int_{0}^{\rho }\chi _{T}^{-1}d\rho \ .$ When applied to the first term on
the r.h.s. of Eq. ($\ref{b3}$), $\left( \partial \beta P/\partial \rho
\right) _{T,{\bf x}}^{{\rm HS-PY}c}$, this procedure does indeed lead to the
known PY$_{c}$ - EOS for HS mixtures. For the SHS models, the integration
requires the knowledge of the dependence (if any) of the coefficients $%
K_{ij} $ on density.

In the SHS1-PY case, the aforesaid calculation is nevertheless practically
impossible, due to the lack of an explicit expression for $\lambda
_{ij}(\rho )$. This difficulty has been by-passed by exploiting a further
result by Baxter, which directly provides the PY pressure itself rather than
the\ inverse susceptibility \cite{Perram75,Barboy79}. On the other hand, in
the SHS2-MSA case the density-independence of $K_{ij}^{{\rm SHS2-MSA}}$
might suggest that the integration of $\chi _{T}^{-1}$ is a straightforward
operation. Unfortunately, this is not so.

In fact, to ensure that the whole procedure is correct, one must first test
whether the differential $\sum_{i}\left( \partial \beta P/\partial \rho
_{i}\right) d\rho _{i},$ constructed with the partial derivatives given by
Eq. ($\ref{b1}$), is exact, since $P$ must be a state function. For this to
occur, it is necessary that

\begin{equation}
\frac{\partial }{\partial \rho _{j}}\left( \frac{\partial \beta P}{\partial
\rho _{i}}\right) =\frac{\partial }{\partial \rho _{i}}\left( \frac{\partial
\beta P}{\partial \rho _{j}}\right) ,  \label{b4}
\end{equation}%
for any pair $i$ and $j.$ This symmetry condition, obeyed by any exact
theory, may not be met when using an approximate closure. In this case, no
compressibility EOS can exist within the considered approximate theory. Now,
we show that the equality ($\ref{b4}$) of the mixed second-order partial
derivatives is not necessarily satisfied for SHS mixtures.

Let us assume, rather generically, that a certain SHS model with an
appropriate closure has a solution $q_{ij}(r)$ given by Eqs. ($\ref{s2}$)-($%
\ref{s4}$) with {\it density-independent}, symmetric,{\it \ }coefficients $%
M_{ij}=12K_{ij}=M_{ji}.$ Inserting this solution into Eq. ($\ref{b1}$) and
taking the derivative of $\ \partial \beta P/\partial \rho _{i}$ with
respect to $\rho _{j}$ yields a rather lengthy expression for $\partial
^{2}\beta P/\partial \rho _{j}\partial \rho _{i}.$ Upon discarding all terms
which are apparently symmetric with respect to an exchange of indices $i$
and $j$, we are left with the following sum 
\begin{eqnarray}
&&S_{1}(ij)\equiv -\frac{1}{\Delta }\sigma _{i}M_{ij}X_{j}+\frac{6\xi _{2}}{%
\Delta ^{3}}\sigma _{i}^{3}\sigma _{j}X_{j}  \nonumber \\
&&+\frac{1}{\Delta ^{2}}\left[ \sigma _{i}^{3}\left( X_{j}-X_{j}^{2}\right)
+3\sigma _{i}^{2}\sigma _{j}X_{j}+\sigma _{i}^{3}\sigma _{j}\left(
X_{j}^{(0)}-\frac{\pi }{6}\sum_{l}\rho _{l}X_{l}M_{jl}\right) \right] ,
\label{b5}
\end{eqnarray}%
where $X_{i}^{(0)}\equiv (\pi /6)\sum_{l}\rho _{l}M_{il}.$ A compressibility
EOS can exist, within an approximate theory of the considered kind, only if
the relevant coefficients $M_{ij}$ are such that $S_{1}(ji)=S_{1}(ij).$

Tutschka and Kahl's choice \cite{Tutschka98,Tutschka01} does not meet this
requirement. On the other hand, it is easy to verify that even {\it any}
choice with {\it factorized }coefficients, $M_{ij}=M_{0}G_{i}G_{j},$\ fails
to satisfy the necessary condition ($M_{0}$ is a {\it density-independent}
factor, which in our model defined by Eqs. ($\ref{s6}$)-($\ref{s7}$)
coincides with $1/T^{\ast }$). $S_{1}(ij)$ cannot be symmetric for a generic
choice of $G_{m}$'s, but the same occurs if we assume the power-law
relationship between stickiness and size employed in Refs. \cite%
{Gazzillo00,Gazzillo02}, i.e. $G_{m}=\sigma _{m}^{\alpha }\ /\left\langle
\sigma \right\rangle ^{\alpha -1},$ where $\left\langle \sigma \right\rangle 
$ denotes the average diameter and $\alpha \geq 0.$ In this case, the value $%
\alpha =1$ is however indicated as preferable, since it represents the only
way of making the first term, $-\Delta ^{-1}\sigma _{i}M_{ij}X_{j}$,
symmetric.

As a first result, one can thus state that no compressibility EOS can exist
within the SHS2-MSA theory with coefficients given either by any
factorization rule $M_{ij}=M_{0}G_{i}G_{j},$ or by Tutschka and Kahl's
unfactorized choice \cite{Tutschka98,Tutschka01}. We note that in Ginoza's
recent analysis \cite{Ginoza01}, such a crucial feature is failed to be
recognized, and the corresponding expression for the compressibility EOS is
therefore incorrect.

It is clear that the violation of the relation ($\ref{b4}$) can be traced
back to the inadequacy of the MSA closure, which is responsible for the
density-independence of the coefficients $M_{ij}$ (or $K_{ij}$). More
generally, one might suspect that no compressibility EOS can exist for any
solution $q_{ij}\left( r\right) $ given by Eqs. ($\ref{s2}$)-($\ref{s4}$) as
long as the $M_{ij}$'s are density-independent.

Let us now assume that the $M_{ij}$'s depend on the densities $(\rho
_{1},\ldots ,\rho _{p})$. In this case, $\partial ^{2}\beta P/\partial \rho
_{j}\partial \rho _{i}$ contains further contributions stemming from
derivatives. Again discarding those that are clearly symmetric, one gets the
following terms

\begin{eqnarray}
S_{2}(ij) &\equiv &\frac{1}{\Delta }\sum_{l}\rho _{l}\left( \sigma _{i}+%
\frac{3\xi _{2}}{\Delta }\sigma _{i}\sigma _{l}+\sigma _{l}\right) \frac{%
\partial M_{il}}{\partial \rho _{j}}\   \nonumber \\
&&+\frac{1}{\Delta ^{2}}\ \frac{\pi }{6}\sum_{l,m}\rho _{l}\rho _{m}\left(
2\sigma _{i}^{3}+\frac{6\xi _{2}}{\Delta }\sigma _{i}^{3}\sigma _{l}+3\sigma
_{i}^{2}\sigma _{l}\right) \sigma _{m}\frac{\partial M_{ml}}{\partial \rho
_{j}}  \nonumber \\
&&-\frac{1}{\Delta }\sigma _{i}\ \frac{\pi }{6}\sum_{l,m}\rho _{l}\rho _{m}\
\sigma _{m}\frac{\partial \left( M_{il}M_{ml}\right) }{\partial \rho _{j}} 
\nonumber \\
&&-\frac{2}{\Delta ^{2}}\sigma _{i}^{3}\left( \frac{\pi }{6}\right)
^{2}\sum_{l,m,n}\rho _{l}\rho _{m}\rho _{n}\ \sigma _{m}\sigma _{n}\frac{%
\partial \left( M_{ml}M_{nl}\right) }{\partial \rho _{j}},  \label{b6}
\end{eqnarray}%
which must be added to $S_{1}(ij),$ to form a new symmetry condition for $%
S(ij)\equiv S_{1}(ij)+S_{2}(ij).$

We emphasize the fact that the density-dependent coefficients of the SHS1-PY
solution, as given by Eq. ($\ref{s9}$), should satisfy the requirement $%
S(ji)=S(ij)$, but a direct test of this feature proves to be a highly
non-trivial task. However, this observation prompts the suggestion that a
simple density-dependent modification of the best SHS2-MSA choice, $%
M_{ij}=M_{0}\sigma _{i}\sigma _{j},$ along the lines of the SHS1-PY
solution, might lead to a fulfilment of the symmetry condition. Indeed, we
find that the choice

\begin{equation}
M_{ij}=M_{0}\frac{1}{\Delta }\sigma _{i}\sigma _{j}\ ,\qquad \text{with}%
\qquad M_{0}=\phi \left( \frac{1}{T^{\ast }}\right) ,  \label{b7}
\end{equation}%
(where $\phi $ is an arbitrary function vanishing as $T^{\ast }\rightarrow
\infty $) leads to a compressibility pressure $P$ satisfying the condition ($%
\ref{b4}$).

Next, we discuss a possible physical origin of the factor $1/\Delta $
appearing in the above solution. First, we note that in $K_{ij}^{{\rm SHS1-PY%
}}$ of Eq. (\ref{s9}) one could relate $\sigma _{ij}^{2}$ to a measure of 
{\it adhesive interaction surface} (since the centre of a particle $i$ which
moves around a particle $j,$ but remains in contact with it, must lie on a
spherical surface with radius $\sigma _{ij}$), while the density-dependent
factor $y_{ij}(\sigma _{ij})$ represents the probability of finding a
particle of species $i$ touching any given particle of species $j$. On the
other hand, in the corresponding {\it modified} expression (\ref{b7}), one
could imagine $\sigma _{i}\sigma _{j}$ as representing the area of an
interaction spherical surface with radius $\left( \sigma _{i}\sigma
_{j}\right) ^{1/2},$ while $1/\Delta $ may be reckoned as a crude
approximation to $y_{ij}(\sigma _{ij})$ ($1/\Delta $ is indeed the {\it %
simplest} term, independent of the species indices $i$ and $j$, appearing in
the expression of cavity functions at contact).

There are (at least) two ways to relate the result (\ref{b7}) to some
possible closure, which should represent an improvement over the MSA one.
First, one might consider a {\it generalized mean spherical approximation}
(GMSA) \cite{Caccamo96} instead of the MSA. This amounts to replace the MSA, 
$c_{ij}\left( r\right) =-\beta u_{ij}^{{\rm HSY}}(r)=z\beta
A_{ij}e^{-z(r-\sigma _{ij})}/r$ for $r>\sigma _{ij},$ with the Yukawa
closure $c_{ij}\left( r\right) =zK_{ij}e^{-z(r-\sigma _{ij})}/r$ for $%
r>\sigma _{ij}$, with parameters $K_{ij}$ not given by Eq. (\ref{s5}), but
density-dependent and initially undetermined. The GMSA has often been used
in the past, and its unknown coefficients $K_{ij}$ have usually been
determined by employing some thermodynamic consistency condition. In the
present case, one could regard the symmetry condition (\ref{b4}) as an
alternative condition for the GMSA closure. Within this conceptual
framework, Eq. (\ref{b7}) is the simplest solution.

Alternatively, we recall that the PY approximation may be written in the
form $c_{ij}\left( r\right) =f_{ij}\left( r\right) y_{ij}\left( r\right) ,$
while the MSA may be derived from it under the approximation $f_{ij}\left(
r\right) \simeq -\beta u_{ij}(r)$, with $y_{ij}\left( r\right) \simeq 1,$
for $r>\sigma _{ij}.$ Hence one could define a new density-dependent closure
for SHS potentials, assuming $c_{ij}\left( r\right) =-y_{ij}(\sigma _{ij})\
\beta u_{ij}(r)$ for $r>\sigma _{ij}.$ The use of $u_{ij}^{{\rm HSY}}(r)$
with factorized coefficients, along with the rough approximation $%
y_{ij}(\sigma _{ij})\simeq 1/\Delta ,$ would then lead to the $M_{ij}$
coefficients expressed by Eq. (\ref{b7}),$\ $with $\phi \left( 1/T^{\ast
}\right) =1/T^{\ast }.$

Finally, it is worth reporting that the solution corresponding to Eq. (\ref%
{b7}) yields the following compressibility EOS

\begin{equation}
\beta P^{c}\ {\cal V}=\frac{\eta }{1-\eta }+\left( 3-M_{0}\right)
e_{1}\left( \frac{\eta }{1-\eta }\right) ^{2}\ +\frac{1}{3}\left(
3-M_{0}\right) ^{2}e_{2}\left( \frac{\eta }{1-\eta }\right) ^{3},  \label{b8}
\end{equation}

\noindent where ${\cal V}=(\pi /6)\left\langle \sigma ^{3}\right\rangle $
denotes the average volume of a particle,  $\eta \equiv \xi _{3}=\rho {\cal V%
}$ is the packing fraction, while the dimensionless parameters $e_{1}\equiv
\left\langle \sigma \right\rangle \left\langle \sigma ^{2}\right\rangle
/\left\langle \sigma ^{3}\right\rangle $, $\ e_{2}\equiv \left\langle \sigma
^{2}\right\rangle ^{3}/\left\langle \sigma ^{3}\right\rangle ^{2}$ depend on
the molar composition ${\bf x}$ and reduce to $e_{1}=e_{2}=1$ for
one-components fluids. Note that $M_{0}=$ $\phi \left( 1/T^{\ast }\right) $
may or may not coincide with $1/T^{\ast }$, but it must vanish as $T^{\ast
}\rightarrow \infty $, so that the PY$c$ - EOS for HS mixtures is recovered.

The simplicity of Eq. (\ref{b8}) is due to the fact that now $X_{i}=M_{0}\xi
_{2}\sigma _{i}/\Delta ,$ and thus we get

\begin{equation}
a_{i}=\frac{1}{\Delta }+\left( 3-M_{0}\right) \frac{\xi _{2}\sigma _{i}}{%
\Delta ^{2}},\qquad b_{i}=-\left( 3-M_{0}\right) \frac{\xi _{2}\sigma
_{i}^{2}}{2\Delta ^{2}},  \label{b9}
\end{equation}%
which differ from their HS counterparts only in having the HS factor 3
replaced with the temperature-dependent coefficient $3-M_{0}.$

In spite of its plainess, Eq.(\ref{b8}) represents an analytical and 
consistent compressibility EOS for a multicomponent system of particles
with both repulsive and attractive interactions. Furthermore, it
bears interesting connections with the SHS1-PY model which will
be discussed elsewhere.

\section{CONCLUSIONS}

The PY compressibility EOS of Baxter's SHS1 model is practically
inapplicable to fluids with a large number of components. This difficulty
urges to search for either alternative closures or different models.

In this paper we have investigated the SHS2 model, which is analytically
solvable within the MSA, and whose thermodynamics is still rather
unexplored, especially in the multi-component case. 
In particular we have focused on the 
compressibility EOS, since it is the simplest route to get the pressure from $%
q_{ij}\left( r\right) $.

Mixtures require a more careful analysis than pure fluids, and one should
always consider the possibility that an
approximate theory suffers from some thermodynamic inconsistencies not present
in the one-component counterpart. In particular, in order
to ensure the existence of \ the compressibility EOS, we have pointed
out the necessity of an explicit check of a basic thermodynamic consistency
requirement hinging on the equality of the mixed
second-order partial derivatives of $P$ with respect to the densities, which
, in turn, can be expressed in terms of $q_{ij}\left( r\right) .$ 
This crucial feature is often overlooked in the literature, perhaps
because it would be automatically fulfilled in an {\it exact} theory. In
an approximate theory, on the other hand, this is not the case, and 
the reciprocity condition ($\ref{b4}$) is one of the sum-rules that
must be explicitly checked for. 

Two are the main results of the present work.
A first finding is that, although an MSA compressibility EOS is known for
pure SHS2 fluids \cite{Mier89}, its extension to mixture is not possible,
since the SHS2-MSA solution violates the aforesaid symmetry
condition. As a direct consequence, the compressibility EOS 
recently reported by
Ginoza \cite{Ginoza01} is flawed. We have discussed how this inconsistency 
occurs
for any choice of factorized stickiness coefficients, as well as for
Tutschka and Kahl's unfactorized ones \cite{Tutschka98,Tutschka01}, and 
argued that it has its origin in the deficiency of the MSA closure.
The reason appears to be the density-independence of
the MSA coefficients $K_{ij}=$ $q_{ij}(\sigma _{ij}^{-}).$
Second, we have presented an illustrative example, where the inclusion
of a simple and plausible
density-dependence in the matrix $K_{ij},$ produces a fulfilment
of the required condition, and generates a possible compressibility EOS. 

This result seems to suggest that no compressibility EOS can
exist for any closure leading to a solution $q_{ij}\left( r\right) $
of the form given by Eqs. ($\ref{s2}$)-($\ref{s4}$) 
with density-independent $K_{ij}$
coefficients. This also prompts the necessity of abandoning the MSA
and resort to density-dependent closures. We have attempted to do this
by interpreting our result for the compressibility EOS, as originating
from a different, more sophisticated, approximate theory. 

Although it is clear that our result cannot be considered  as
the correct final solution to our problem, it nevertheless represents,
not a simple academic exercise, but a useful
step towards a satisfactory EOS for SHS mixtures.

\end{document}